\documentclass[11pt,a4paper]{article}%
\usepackage{amsmath,hyperref,url}
\usepackage{natbib}
\usepackage{amsfonts}
\usepackage{amssymb}
\usepackage{graphicx}
\usepackage{booktabs}
\usepackage[english]{babel}
\providecommand{\U}[1]{\protect\rule{.1in}{.1in}}
\usepackage{setspace}
\begin{document}

\title{Central Clearing of OTC Derivatives:\\bilateral vs multilateral netting
\footnote{Presented at the Global Derivatives Conference 2011, the 5th Financial Risks Forum: Systemic Risk, the European Institue on Financial Regulation (Paris, 2011), the SIAM Conference on Financial Mathematics \& Engineering 2012 and the Midwest Finance Association 2013 Annual Meeting. We  thank the Europlace Institute of Finance and the Danish Council for Independent Research $|$ Social Sciences for financial support. }\bigskip
}
\author{Rama Cont\smallskip\\{\small{Imperial College London}}\\
\small{\tt Rama.Cont@imperial.ac.uk}
\and Thomas Kokholm\smallskip\\
{\small {Aarhus University, Denmark}} 
\\{\small \tt{  thko@asb.dk}}}
\date{2012}
\maketitle
\begin{abstract}
\thispagestyle{empty}
We study the impact of central clearing of over-the-counter (OTC) transactions on
counterparty exposures in a market with OTC transactions across several
asset classes with heterogeneous characteristics.
The impact  of
introducing a central counterparty (CCP)  on  expected interdealer exposure is determined by the tradeoff between  multilateral netting across dealers on one hand and  bilateral netting across asset classes on the other hand. We find this tradeoff  to be 
sensitive to assumptions on heterogeneity of asset
classes in terms of `riskyness' of the asset class as well as
correlation of exposures across asset classes.
In particular, while an analysis assuming independent, 
homogeneous exposures suggests that central clearing is efficient only if  one has an unrealistically high
number of participants,  the opposite conclusion is reached if
differences in riskyness and correlation across asset classes are realistically taken into account. 
We argue that empirically plausible specifications of model parameters lead to the conclusion that central clearing does reduce interdealer exposures: the
gain from multilateral netting in a CCP overweighs the loss of netting across asset classes in bilateral netting agreements.
When a CCP exists for interest rate derivatives, adding a CCP for credit
derivatives is shown to decrease overall exposures.
These findings are shown to be robust to the statistical assumptions of the model as well as the choice of risk measure used to quantify exposures.
\end{abstract}

\newpage

\tableofcontents

\thispagestyle{empty}

\newpage

\section{Central clearing of OTC derivatives}\label{OTCmarkets}
\setcounter{page}{1}

\begin{figure}[h]
  \begin{center}
    \includegraphics[width=0.9\textwidth]{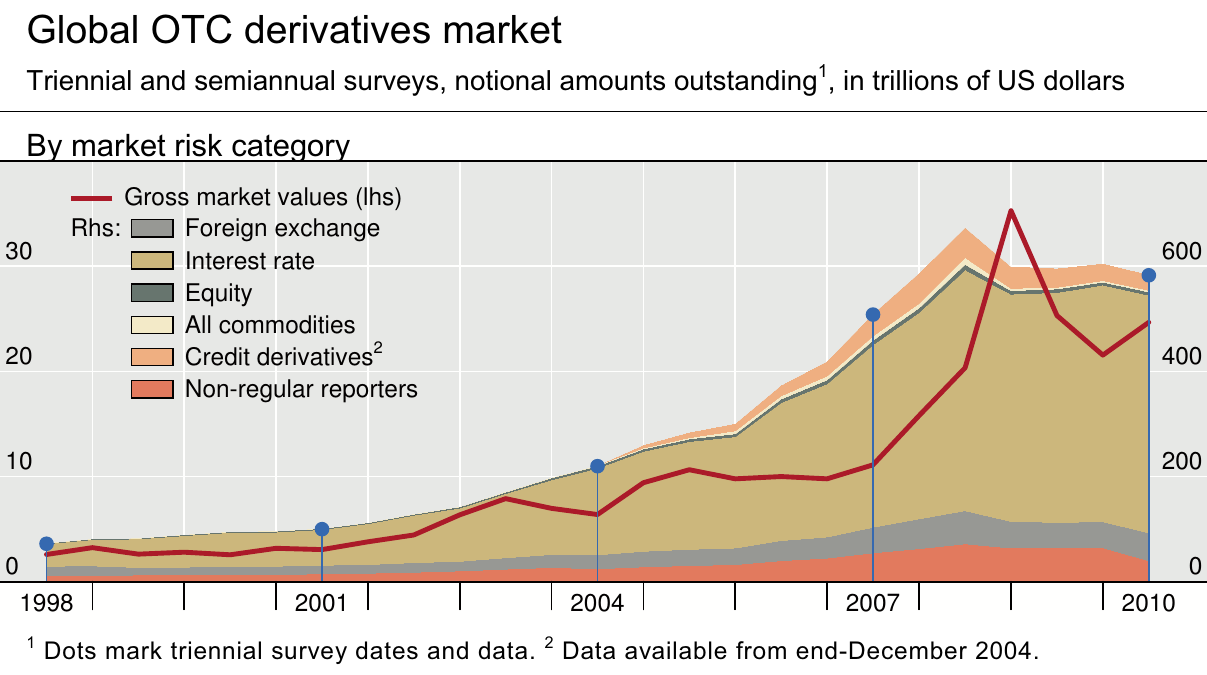}
    \caption{Notional amounts outstanding in  OTC derivatives markets, in trillions USD. Source: BIS.}
    \label{Notionals_fig}
  \end{center}
\end{figure}

Over-The-Counter (OTC) derivatives represent a sizable fraction of  financial transactions worldwide, encompassing a wide variety of contracts and asset classes.
Figure \ref{Notionals_fig} depicts the development in the OTC derivatives markets for different asset
classes since 1998. Although the increase in notionals has stopped since the peak of the financial
crisis the overall growth is impressive. The contraction in some OTC derivatives asset classes since
the beginning of the crisis is clearly seen from Table \ref{notionals.table}, which shows gross notional
values in different asset classes as of June 2007 and June 2010.

\begin{table}[h] \centering
\caption{Gross notional values in OTC derivatives markets in billions as of June 2007 and June 2010. Source: BIS}\smallskip
\begin{tabular}
[c]{lrr}\toprule
 Asset Class & 2007 & 2010 \\\midrule
 Commodity & 8,255 & 3,273\\
 Equity Linked & 9,518 & 6,868\\
 Foreign Exchange & 57,604 & 62,933\\
 Interest Rate & 381,357 & 478,093\\
 Credit Derivatives & 51,095 & 31,416\\
 Other & 78 & 72\\\midrule
 Total & 507,907 & 582,655\\\bottomrule
\end{tabular}\label{notionals.table}
\end{table}

Most bilateral OTC contracts involve exchange of collateral between counterparties, to mitigate counterparty risk. 
However, there is  evidence that collateral requirements in bilateral Credit Support Annexes (CSAs) lead to non-negligible residual counterparty exposures.
This is particularly the case in credit derivatives contracts where jump-to-default risk is typically not covered by bilateral collateral requirements \citep{cont2010credit}. In fact, the collapse of AIG  in September 2008 was triggered by CDS 
collateral requirements it could not meet.
\cite{brigo2011arbitrage} and \cite{fujii2011cds} both show that even in the extreme case of continuously
collateralized CDS contracts, counterparty risk is present, e.g. jump-to-default risk.
\cite{arora2012} find evidence of counterparty risk in the interdealer CDS markets. \cite{singh2010collateral} finds that most large banks active in the OTC derivatives market
are under-collateralized. 

One of the solutions proposed for mitigating counterparty risk is the central clearing of OTC derivatives. The Dodd-Frank Wall Street Reform and Consumer Protection Act of 2010 has mandated central
clearing for standardized Over-The-Counter (OTC) derivatives. Many classes of OTC derivatives are already being partially
cleared through Central Counterparties (CCPs), e.g. LCH Clearnet for Interest Rate Swaps (IRSs),
ICE Clear and CME  for Credit Default Swaps (CDS), Fixed Income Clearing Corp (FICC)
for fixed-income OTC derivatives and many others. The volume of OTC derivatives  cleared through central counterparties will most
likely increase with the implementation of the Dodd-Frank Act and increased capital requirements in Basel 3 for non-cleared OTC transactions.

With a CCP, a bilateral OTC derivative trade between two counterparties is replaced by
two symmetric trades between the CCP and each counterparty.
Trading OTC derivatives through a CCP has some desirable effects: in a bilateral market
default of one entity can spread throughout the system leading to a chain of contagious defaults
\citep{cont2010network,iyer2011interbank}. A CCP can break this chain of contagion. It has been argued that clearing trades through a  CCP leads
to a gain in multilateral netting among market participants in the asset class being cleared,
higher transparency, risk sharing among members of the clearinghouse, no duplicative
monitoring, mitigation of counterparty risk as counterparties are insulated from each other's default,
a reduction in commitment frictions \citep{nosal2011clearing}. Moreover, by centralizing information, supervision and
transparency for regulators are facilitated while retaining trade anonymity,
and from a moral hazard perspective it is less problematic to bail out a CCP
compared to bailing out individual banks.
While a CCP leads to higher multilateral netting it comes at the cost of reduced bilateral netting:
if only a subset of OTC contracts is cleared,  netting {\it across } asset classes due to
hedging is not accounted for, e.g. if a fixed income position is hedged with an OTC interest rate derivative,
separate clearing of the derivative leads to larger collateral requirements.


\begin{table}[htbp] \centering
\caption{Notional  OTC derivative exposures for the 10 largest US derivatives dealers, March 31, 2009  (billions USD). Source: Office of the Comptroller of the Currency.}\smallskip
\begin{tabular}
[c]{lrrrr}\toprule
 Dealer & Forwards & Options & Swaps & Credit \\\midrule
 JP Morgan Chase & 8,422 & 10,633 & 51,221 & 7,495\\
 Bank of America & 9,132 & 6,908 & 50,702 & 5,649\\
 Goldman Sachs & 1,631 & 6,754 & 30,958 & 6,601\\
 Morgan Stanley & 1,127 & 3,530 & 26,112 & 6,307\\
 Citigroup & 4,743 & 5,868 & 15,199 & 2,950\\
 Wells Fargo & 1,217 & 543 & 2,748 & 286\\
 HSBC & 595 & 185 & 1,565 & 913\\
 Taunus & 667 & 20 & 162 & 144\\
 Bank of New York & 371 & 304 & 404 & 1\\
 State Street & 571 & 45 & 24 & 0\\\midrule
 Total & 28,476 & 34,792 & 179,094 & 30,348\\\bottomrule
\end{tabular}\label{indivnotionals2009.table}
\end{table}

\begin{table}[htbp] \centering
\caption{Notional amounts of OTC derivatives contracts of the 10 biggest US derivatives dealers as of December 31, 2010 in billions. Source: Office of the Comptroller of the Currency.}\smallskip
\begin{tabular}
[c]{lrrrr}\toprule
 Dealer & Forwards & Options & Swaps & Credit \\\midrule
 JP Morgan Chase & 11,807 & 8,899 & 49,332 & 5,472\\
 Bank of America & 10,287 & 5,848 & 43,482 & 4,367\\
 Citigroup & 6,895 & 7,071 & 28,639 & 2,546\\
 Goldman Sachs & 3,805 & 8,568 & 27,392 & 4,233\\
 Morgan Stanley & 5,459 & 3,855 & 27,162 & 4,648\\
 Wells Fargo & 1,081 & 463 & 1,806 & 93\\
 HSBC & 758 & 127 & 1,901 & 700\\
 Bank of New York & 420 & 367 & 555 & 1\\
 Taunus & 848 & 21 & 199 & 33\\
 State Street & 599 & 76 & 79 & 0\\\midrule
 Total & 41,959 & 35,295 & 180,547 & 22,093\\\bottomrule
\end{tabular}\label{indivnotionals2010.table}
\end{table}

Tables \ref{indivnotionals2009.table} and \ref{indivnotionals2010.table} show the OTC derivatives
notionals in four different asset classes of the 10 largest US derivatives dealers.
Inspection of the tables reveals a decrease in the notional of the credit derivatives class between 2009 and 2010.
This is primarily  a consequence of increased trade compression and
the increased use of central clearing in the CDS market \citep{vause2010counterparty}:
the first  CDS clearing facility, operated by ICE Clear, started its operations in 2009 and currently clears a sizable fraction of  interdealer  trades in index and single name corporate CDSs. Comparing  Table \ref{indivnotionals2010.table} with Table \ref{indivnotionals2009.table} suggests that the central clearing of CDSs initiated 
in 2009 and the resulting compression of bilateral trades led to  an overall  decrease of the magnitude of dealer exposures.

\begin{figure}[h]
  \begin{center}
    \includegraphics[width=0.45\textwidth]{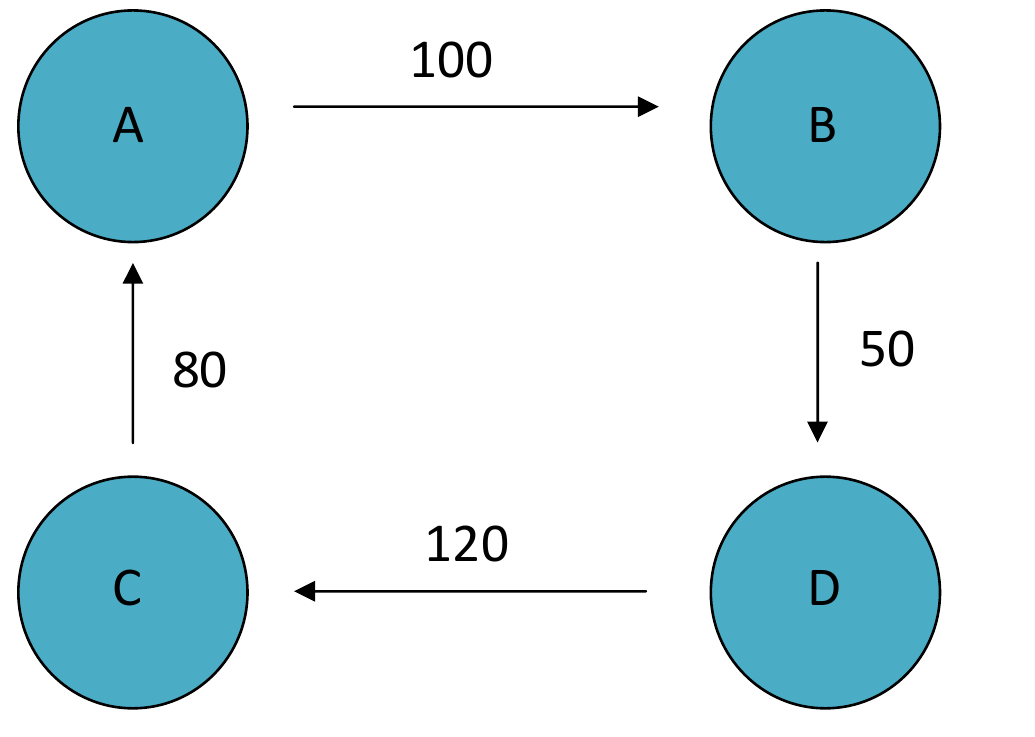}
    \includegraphics[width=0.45\textwidth]{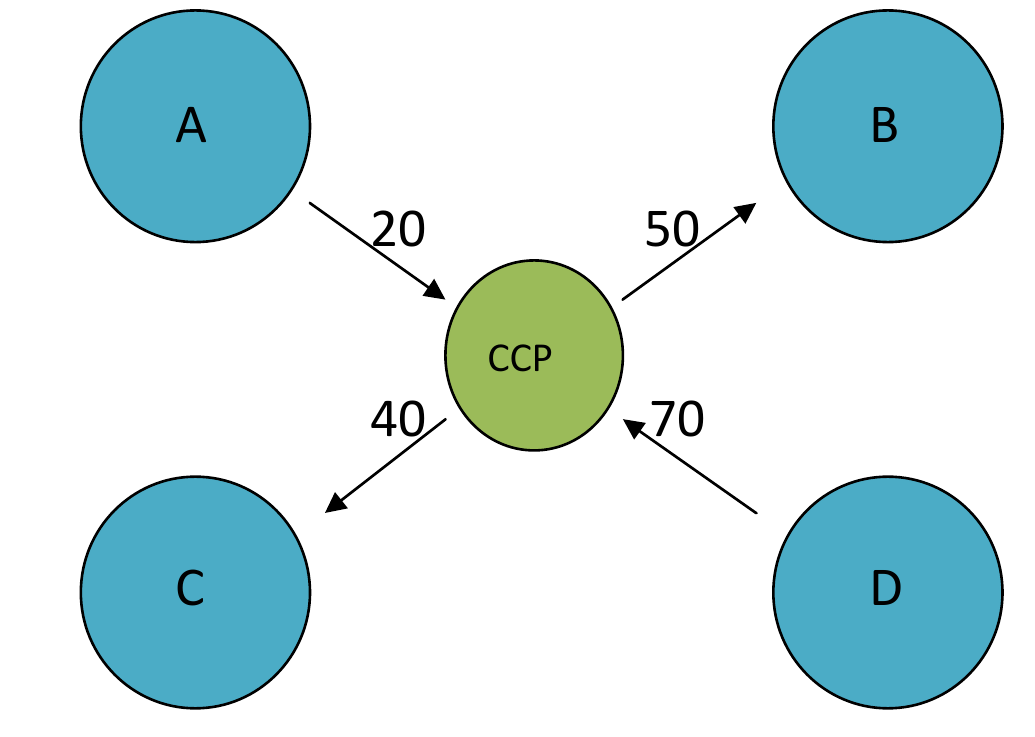}
    \caption{Left: A stylized bilateral market. Right: A centrally cleared market.}
    \label{bilateral_fig}
  \end{center}
\end{figure}

To understand why central clearing of OTC trades can lead to such a decrease in the sum of bilateral exposures,  
consider a stylized market of four participants
with bilateral exposures like shown in the left part of
Figure \ref{bilateral_fig}. In this market the
sum of bilateral exposures amounts to 350, taking into account bilateral netting. 
Introducing a  CCP enables {\it multilateral netting}
of the exposures (Figure \ref{bilateral_fig} , right) which reduces the total net exposure to 180. 

The above example highlights the benefits of multilateral netting through a CCP. But there are also situations where central clearing of a single asset class may actually increase overall net exposures. If only a
subset of OTC contracts is cleared, netting \emph{across} asset classes due to hedging is
not accounted for. For instance, if a fixed income position is hedged with an OTC interest rate derivative,
separate clearing of the derivative leads to larger exposure, which in return leads to larger collateral
requirements. A relevant question is thus to compare the impact of these two competing effects:
\begin{itemize}
  \item the reduction of net exposures due to multilateral netting across counterparties through central clearing, and
  \item the increase in net bilateral exposures through the loss of netting across asset classes.
\end{itemize}
In an influential paper widely disseminated to regulators, \cite{duffiedoes} argue that if only a subset of OTC contracts is cleared
in a clearinghouse, then there may be a loss of netting {\it across} asset classes
which exceeds the gain in multilateral netting across market participants.
This conclusion has been  interpreted by some to mean that 'CCPs
increase net exposure'. 

\paragraph{Contribution}

In the present work, we reconsider the impact of central clearing of over-the-counter (OTC) transactions on
counterparty exposures in a market with OTC transactions across several
asset classes with heterogeneous characteristics.
We find that the impact, on total expected interdealer exposure,  of
introducing a central counterparty for a single class of OTC
derivatives is highly sensitive to assumptions on heterogeneity of asset
classes in terms of `riskyness' of the asset class as well as
correlation of exposures across asset classes.
In particular, while  a model with independent, 
homogeneous exposures suggests that central clearing is efficient only
if  one has an unrealistically high
number of participants,  these conclusions are not robust to departures from the assumptions of independent and homogeneity. In fact, we find that the opposite conclusion is reached if
differences in riskyness across asset classes are realistically taken into account.
Empirically plausible specifications of these parameters lead to the
conclusion that the gain from multilateral netting in a CCP
overweighs the loss of netting across asset classes in bilateral netting agreements.
When a CCP exists for interest rate derivatives, adding a CCP for credit
derivatives is shown to decrease overall exposures. This result distinguishes our
study from \cite{duffiedoes} where the
marginal effect of clearing credit derivatives is very limited when interest rate swaps are
already being cleared.
These findings are shown to be robust to the choice of distribution for OTC credit
derivatives exposures.

\paragraph{Outline}  The remainder of the paper is structured as follows.

Section \ref{Model_section} introduced a stylized model of an OTC market with $K$ different asset classes and examines how the separate or joint clearing of one or more asset classes in a CCP affects bilateral exposures. Section \ref{sec:heterogeneity} emphasizes the key role played by assumptions on heterogeneity of various parameters --heterogeneity of riskyness across asset classes, heterogeneity of exposure sizes across dealers-- and proposed a flexible parameterization to account for this heterogeneity.

 Section \ref{Analysis_sec} uses this framework to study how the introduction
of CCPs affects OTC derivatives exposures, using three different risk measures.
Finally, Section \ref{conclusion_sec} concludes.

\newpage
\section{A stylized model of OTC clearing}\label{Model_section}

To compare the impact on exposures of multilateral netting across counterparties, as opposed to cross-asset netting in OTC markets with bilateral transactions,  we 
consider a market where $N$ participants trade over-the-counter derivatives in $K$ different asset classes, labeled $k=1...K$. 
Collateral requirements, whether bilateral or in CCPs,  are typically based on an estimation of exposures over a horizon which 
we denote by $X_{ij}^k$; the net value
of the total positions at some future time point that $i$ owns of derivatives belonging to asset
class $k$ with $j$ as the counterparty.
The exposure of $i$ to $j$ in asset class $k$ is then defined as
\begin{equation}
\max \left\{ X_{ij}^k , 0 \right\}.
\end{equation} 
We want to compare total exposures across three configurations:
\begin{enumerate}
\item The total exposure $e_i^0$ of company $i$ under
bilateral netting (no central clearing):
\begin{equation}
e_i^0=\sum_{j\neq i} \max \left\{ \sum_{k=1}^K X_{ij}^k , 0
\right\}. \label{bilat_eq}
\end{equation}

\item The exposure $e_i^1$ assuming  asset
class $K$ is cleared through a central counterparty
\begin{equation}
e_i^1=\sum_{j\neq i} \max \left\{  \sum_{k=1}^{K} \left( 1 -
w_{k} \right) X_{ij}^k , 0 \right\} + \underbrace{\max \left\{
\sum_{j\neq i} w_K X_{ij}^K ,0 \right\}}_{\rm Exposure\ to\ CCP},
\label{central_eq}
\end{equation}
where $w_k$ denotes the fraction of asset class $k$ that is being cleared. Hence in this case we
have that, $w_k=0$ for $k=1,...,K-1$.
\item The exposure $e_i^2$ when two asset
classes $K-1$ and $K$ are separately cleared through a central counterparty
\begin{align}
e_i^2 = & \sum_{j\neq i} \max \left\{ \sum_{k=1}^{K} \left( 1
- w_{k} \right) X_{ij}^k , 0 \right\}
+ \underbrace{\max \left\{ \sum_{j\neq i} w_{K-1} X_{ij}^{K-1} ,0 \right\}}_{\rm Exposure\ to\ CCP\ 1}\nonumber\\
& + \underbrace{\max \left\{ \sum_{j\neq i} w_K X_{ij}^K ,0 \right\}}_{\rm Exposure\ to\ CCP\ 2},
\label{central2ccp_eq}
\end{align}
where $w_k=0$ for $k=1,...,K-2$.
\end{enumerate}

In equation \eqref{bilat_eq} all netting is done bilaterally across all $K$ asset classes and the total net
exposure of dealer $i$ is the sum of the net exposure with each counterparty $j$. When a CCP is
introduced in the market, as in equations \eqref{central_eq}
and \eqref{central2ccp_eq}, the total net exposure is the sum of
the bilateral net exposures to all counterparties and the exposures to the CCPs.
Hence, the introduction of clearinghouses transfers exposure reductions from bilateral
netting to multilateral netting. Whether the introduction of a clearinghouse
increases or decreases net exposures depends on the particular market, e.g. the notional sizes of the
asset classes, riskyness of the asset classes, correlation between the asset classes, the number of asset classes, the number of banks etc.
In terms of dealer risk management our analysis underestimates the risk reduction from central clearing
since we put the same risk weights on bilateral dealer exposures and
exposures to a CCP. In most cases the riskyness
of a CCP will be lower than that of a dealer - and hence have a smaller risk weight.
The reader is referred
to \cite{arnsdorf2012quantification} for a model to quantify the CCP risk
a dealer faces when participating in a CCP.

In the sequel we consider different distributional choices for the exposures,
and we compare expected net exposures, Value at Risk (VaR), expected shortfall and the
mean of the maximum of realized exposures across dealers across various clearing scenarios.

\section{Accounting for heterogeneity}\label{sec:heterogeneity}

\subsection{The importance  of correlation and heterogeneity in exposures}

The stylized model above requires assumptions on the joint distribution of exposures.

Let us first consider the baseline case where $X_{ij}^k \sim N\left(0,\sigma_k \right)$ are
normally distributed for all $i\neq j$ and allowed to be correlated across asset classes.
As in  \cite{duffiedoes}, we assume that the standard deviation of $X_{ij}^k$ is proportional
to the credit exposure
\begin{equation}
\sigma_k=\alpha_k CE_k.
\end{equation}

\begin{table}[htbp] \centering
\caption{Gross market values in OTC derivatives markets in billions as of June 2010. Source: BIS}\smallskip
\begin{tabular}
[c]{lr}\toprule
 Asset Class & Exposure \\\midrule
 Commodity & 457 \\
 Equity Linked & 706 \\
 Foreign Exchange & 2,524 \\
 Interest Rate & 17,533 \\
 CDS & 1,666 \\
 Other & 1,788 \\\midrule
 Total & 24,673 \\\bottomrule
\end{tabular}\label{exposures.table}
\end{table}

Table \ref{exposures.table} shows
the gross credit exposures $CE_k$ in six different asset classes as of June 2010.
OTC interest rate  derivatives clearly account for
the major portion of this value.

\begin{figure}[h]
  \begin{center}
    \includegraphics[width=0.9\textwidth]{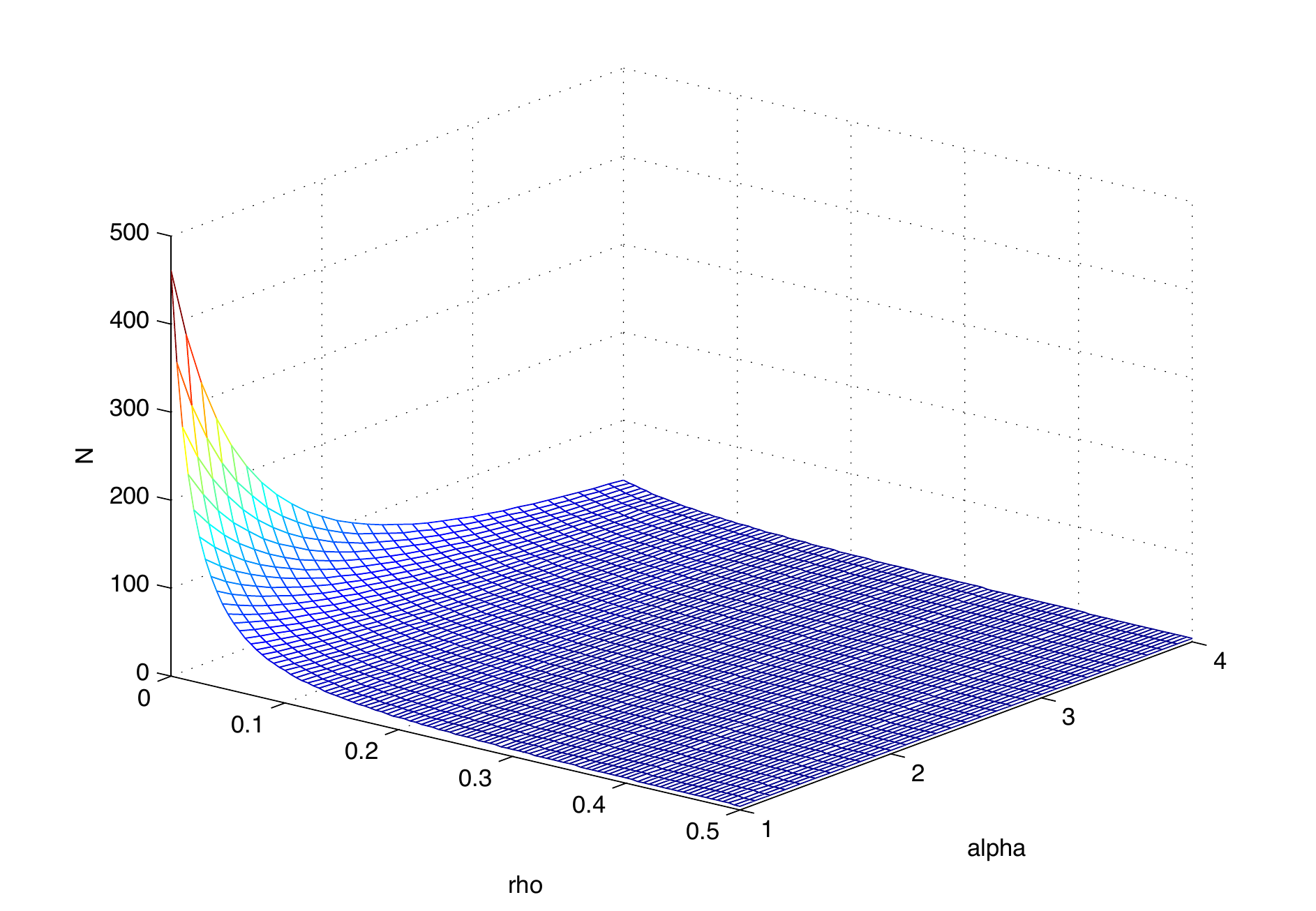}
    \caption{Minimal number of members  $N$ required to participate in the CCP in order to see a reduction in the expected
    exposures as a function of riskyness of the asset class being cleared $\alpha_K$ and the correlation between asset classes $\rho$.}
    \label{ClearingMembers_fig}
  \end{center}
\end{figure}

Consider first a market where only one single asset class is being cleared.
In \cite{duffiedoes} it is shown that if $\alpha_k=\alpha$ for all $k=1,...,K$ and if the
exposures are uncorrelated, the number of entities required
to participate in the central clearinghouse has to be equal to or greater
than 461 in order to
see a reduction in expected exposures: this is an unrealistically high number, so it would imply in practice that existing clearing schemes do not result in reduction of exposures.
But, altering the  assumption on risk per dollar notional to e.g. $\alpha _K=3\alpha $ in the same model, the picture is
changed dramatically and the threshold number drops to 54.
If furthermore we assume the  correlation between the asset class exposures is, e.g., $10\%$, then the number of clearing members needed to achieve compression of exposures is down to... 17!
Table \ref{participants.table} reports the required number of participants for different scenarios.
Figure \ref{ClearingMembers_fig} depicts the minimal number of members required to participate in
the clearinghouse in order to see expected exposure reductions as a
function of correlation and riskyness of the cleared asset class. The Figure clearly reveals that the base case considered in \cite{duffiedoes}. $\rho=0$ and $\alpha_K=1$, is a singularity: a slight deviation from these parameters alters the conclusions of the model dramatically.  

\begin{table}[htbp] \centering
\caption{Minimal number of clearing members required in order for a
CCP to reduce expected exposure. $X^k_{ij}\sim N(0,\alpha_k^2CE_k^2)$ are
jointly Gaussian.}\smallskip
\begin{tabular}
[c]{lr}\toprule
 Assumptions &  $N$\\\midrule
  Duffie \& Zhu's example: &   \\
 $\rho=0,\alpha_i=\alpha$ & 461  \\
 \midrule
 $\rho=0,\alpha_K=3\alpha$   &  \\
  $ \alpha_i=\alpha, i<K$   & 54 \\
  \midrule
  $\rho=0.1,\alpha_K=3\alpha$   &  \\
  $ \alpha_i=\alpha, i<K$   & 17 \\
  \midrule
  $\rho=0.2,\alpha_K=2\alpha$   &  \\
  $ \alpha_i=\alpha, i<K$   & 11 \\
  \bottomrule
\end{tabular}\label{participants.table}
\end{table}

These observations illustrate that, even in a simple setting with (Gaussian) identically distributed  exposures, the tradeoff
between netting across asset classes versus multilateral netting across market
participants is highly sensitive to  assumptions on the  correlation of exposures
and the heterogeneity of
riskyness across asset classes. Conclusions based on a model with
homogeneous risk across asset classes are not robust to departures
from this homogeneity assumption.

We now introduce a model which allows to parametrize this heterogeneity of exposure sizes across dealers and asset classes and differentiates various asset classes in terms of their risk characteristics, as measured by the 'risk' (or collateral requirement) per dollar notional.

\subsection{Accounting for heterogeneity across dealers} 

Instead of credit exposures we now use notional values to determine the
distribution of the $X_{ij}^k$s and assume that they are proportional
to the notional values via
\begin{equation}
X_{ij}^k=\beta_k Z_i^k \frac{Z_j^k}{\sum_{h\neq i} Z_h^k} Y^k_{ij},
\label{exposureGauss_eq}
\end{equation}
where $Y^k_{ij}$ is a random variable with zero mean and unit variance, and
$Z_i^k$ is the total notional size of
dealer $i$ in asset class $k$. In this way the exposure of $i$ to $j$ in class $k$
is a weighted fraction of company $i$'s total
notional in derivatives class $k$, where the weight equals $j$'s notional
size in asset class $k$ relative to the total
notional size of the dealers in asset class $k$.

\subsubsection{Normally distributed exposures}

We first study the case where $Y^k_{ij}$ is a standard normally distributed variable.
In the Gaussian case we can get closed formulas for the expected net exposures
\begin{align}
\mathbf{E}\left[e_i^0\right] = & \frac{1}{\sqrt{2\pi}}\sum_{j\neq i} \sqrt{\sum_{k=1}^K \sum_{m=1}^K \rho_{km} \beta_k Z_i^k \frac{Z_j^k}{\sum_{h\neq i} Z_h^k} \beta_m Z_i^m \frac{Z_j^m}{\sum_{h\neq i} Z_h^m}}\\
\mathbf{E}\left[e_i^1\right] = & \frac{1}{ \sqrt{2\pi} }\sum_{j\neq i} \sqrt{\sum_{k=1}^{K} \sum_{m=1}^{K} \left( 1 - w_{k} \right)\left( 1 - w_{m} \right) \rho_{km} \beta_k \frac{ Z_i^kZ_j^k}{\sum_{h\neq i} Z_h^k} \beta_m  \frac{ Z_i^mZ_j^m}{\sum_{h\neq i} Z_h^m}}\nonumber\\
& + \frac{1}{ \sqrt{2\pi}} \beta_K w_K Z_i^K   \frac{\sqrt{ \sum_{j\neq i} \left(Z_j^K \right)^2}}{\sum_{h\neq i} Z_h^K }\\
\mathbf{E}\left[e_i^2\right] = & \frac{1}{ \sqrt{2\pi} }\sum_{j\neq i} \sqrt{\sum_{k=1}^{K} \sum_{m=1}^{K} \left( 1 - w_{k} \right)\left( 1 - w_{m} \right) \rho_{km} \beta_k  \frac{Z_i^k Z_j^k}{\sum_{h\neq i} Z_h^k} \beta_m  \frac{Z_i^m Z_j^m}{\sum_{h\neq i} Z_h^m}}\nonumber\\
& + \frac{1}{ \sqrt{2\pi}} \beta_{K-1} w_{K-1} Z_i^{K-1} \frac{\sqrt{\sum_{j\neq i} \left(Z_j^{K-1} \right)^2}}{\sum_{h\neq i} Z_h^{K-1}} \nonumber\\
& + \frac{1}{ \sqrt{2\pi}} \beta_K w_K Z_i^K   \frac{\sqrt{\sum_{j\neq i} \left(Z_j^K \right)^2}}{\sum_{h\neq i} Z_h^K }
\end{align}
where $\rho_{km}=1$ for $k=m$.

We compute the Value at Risk and expected shortfalls at the 99\% level by simulation
as these quantities are not available in closed form.

\subsubsection{$t$-distributed CDS exposures}

We also assume that the value of the CDS positions are given by
\begin{equation}
X_{ij}^K = \beta_K Z_i^K \frac{Z_j^K}{\sum_{h\neq i} Z_h^K} t^K_{3,ij},
\label{exposuret_eq}
\end{equation}
where $t^K_{3,ij}$ is a normalized $t$-distribution with unit variance and 3 degrees of freedom
\citep{cont2011stat}. The other asset classes are kept normally distributed as in equation \eqref{exposureGauss_eq}.
Correlation between the classes with different distributional assumptions is introduced
via a normal Copula with correlation parameter $\rho$.
In the case of the $t$-distribution the risk measures are not known in analytical closed
form so we will use simulation in order to compute them.

\subsection{Heterogeneity of risk across asset classes}

Another important feature which affects the analysis is the difference of asset classes  in terms of riskyness. For example, an interest rate swap and a CDS with same notional and maturity typically will not lead to the the same collateral requirements, since their risk is not considered to be identical.
A simple metric for this is the 
exposure per dollar notional for a given asset class. 

To obtain an estimate of this parameter for various asset classes, a simple approach is to examine the standard deviation of the historical daily profit/loss, in terms of percentage of contract notional.
For the CDS asset class, we estimate $\beta_K$ as the mean of the standard deviation of the
daily profit-loss
of 5-year credit default swaps on the names constituting the CDX NA IG  HVOL series 12 in the period
July 1st, 2007 to July 1st, 2009. In the estimation the notionals of the CDSs are assumed equal to 1
and the recovery rate equal to 0.4.

For the same period, we compute the $\beta_k$ for the interest rate class as the
standard deviation of the historical
daily profit-loss from holding a 5-year interest rate swap with notional of 1 and we use this
estimate for all asset classes other than the CDS class.

The estimated values are
\begin{align}
\beta_k & = 0.0039 \quad \text{for\ interest\ rate\ swaps} \\
\beta_K & = 0.0098, \quad \text{for\ CDS}  
\end{align}
i.e. the risk per dollar notional is around three times higher for the CDS asset class.

\section{Does central clearing reduce inter-dealer exposures?}\label{Analysis_sec}

Using the setting described in the previous section, we  now  compare the magnitude of inter-dealer exposures under different market clearing scenarios:
\begin{enumerate}
\item[0.] No central clearinghouse.
\item One central clearinghouse for IRSs.
\item One central clearinghouse for CDSs.
\item Two central clearinghouses: one for IRSs and one for CDSs.
\item One central clearinghouse for both IRSs and CDSs.
\end{enumerate}
Given that, in each of these scenarios, the total interdealer exposure is a random variable, we need  a criterion for evaluating the magnitude of exposures.  \cite{duffiedoes} focus on the {\it expected exposures}; to check for robustness of results with respect to this choice we have computed various risk measures for 
the total interdealer exposure in each scenario:
\begin{itemize}
\item expected exposure
\item $99\% $ quantile of the total exposure
\item $99\% $  tail conditional expectation 
\end{itemize}
These risk measures are reported below in terms of the ratio to their value   under the base scenario 
where no clearinghouse exists.

Currently some markets for OTC derivatives are already being cleared.
Following a meeting on January 27th, 2010 at The Federal Reserve New York, 14 major dealers
say in a letter of March 1st, 2010 that they will aim to centrally clear $w_K=85\%$ of new and
historical credit derivatives trades. In interest rate derivatives, dealers are working to
clear $w_{K-1}=90\%$ of new eligible trades.\footnote{The letter is available for download at\\
http://www.newyorkfed.org/markets/otc\_derivatives\_supervisors\_group.html.} We will use these clearing fractions
in equations \eqref{bilat_eq}-\eqref{central2ccp_eq} when
evaluating the above scenarios. In the case where one single CCP clears both IRSs and CDSs the
classes can be considered
to belong to the same class and we use equation \eqref{central_eq}

We use the notional values reported in Table \ref{indivnotionals2009.table} and assume that there exist
10 identical European dealers, hence the total number of participants is 20. This is not a
high number, at the time of writing the number of participants in
SwapClear at LCH Clearnet is 38.

For the case of the Gaussian CDS exposures we use the analytical formulas to compute the
expected exposures and $10^6$ simulated scenarios to compute the other two risk measures. When
$t$-distributed exposures are involved we approximate the three measures using
$10^6$ simulations and correlate
the exposures across asset classes via a Gaussian Copula.
\begin{table}[htbp] \centering
\caption{Reduction of  interdealer exposures in various clearing scenarios, relative to the base scenario without central clearing, based on OCC data for gross  notional sizes (Q1 2009), assuming
 independence of exposures   across asset classes.
Top: normally distributed CDS exposures.  Bottom: $t$-distributed CDS exposures.
}\smallskip

\makebox[\linewidth]{
\footnotesize{
\begin{tabular}
[c]{lrrrrrrrrrrrrrr}\toprule
 & \multicolumn{4}{l}{\underline{Expected Exposure  }} & & \multicolumn{4}{l}{\underline{Value at Risk  }} & & \multicolumn{4}{l}{\underline{Expected Shortfall  }}\\
Scenario  & IRS & CDS & Two & Joint & & IRS & CDS & Two & Joint & & IRS & CDS & Two & Joint\\
 &   CCP & CCP & CCPs & CCP &  & CCP & CCP & CCPs & CCP & & CCP & CCP & CCPs & CCP\\\midrule
 \multicolumn{4}{l}{\underline{Gaussian CDS exposures}}\\
 JP Morgan Chase & 0.72 & 1.03 & 0.65 & 0.57 & & 0.95 & 1.01 & 0.92 & 0.90 & & 0.97 & 1.01 & 0.94 & 0.92\\
 Bank of America & 0.66 & 1.04 & 0.61 & 0.55 & & 0.93 & 1.01 & 0.91 & 0.90 & & 0.95 & 1.01 & 0.93 & 0.92\\
 Goldman Sachs & 0.78 & 1.01 & 0.62 & 0.52 & & 0.97 & 1.00 & 0.90 & 0.89 & & 0.98 & 1.00 & 0.92 & 0.91\\
 Morgan Stanley & 0.80 & 0.99 & 0.58 & 0.47 & & 0.97 & 1.00 & 0.88 & 0.87 & & 0.98 & 1.00 & 0.91 & 0.90\\
 Citigroup & 0.84 & 1.02 & 0.78 & 0.70 & & 0.98 & 1.01 & 0.95 & 0.94 & & 0.99 & 1.00 & 0.97 & 0.95\\
 Wells Fargo & 0.76 & 1.03 & 0.75 & 0.70 & & 0.96 & 1.01 & 0.95 & 0.93 & & 0.97 & 1.01 & 0.96 & 0.95\\
 HSBC & 1.00 & 0.82 & 0.64 & 0.53 & & 1.00 & 0.96 & 0.89 & 0.88 & & 1.00 & 0.97 & 0.91 & 0.91\\
 Taunus & 1.04 & 0.96 & 0.99 & 0.94 & & 1.01 & 0.98 & 0.99 & 0.98 & & 1.01 & 0.99 & 0.99 & 0.98\\
 Bank of New York & 0.95 & 1.00 & 0.95 & 0.95 & & 0.99 & 1.00 & 0.99 & 0.99 & & 0.99 & 1.00 & 0.99 & 0.99\\
 State Street & 1.01 & 1.00 & 1.01 & 1.01 & & 1.00 & 1.00 & 1.00 & 1.00 & & 1.00 & 1.00 & 1.00 & 1.00\\\midrule
 Total & 0.74 & 1.02 & 0.64 & 0.56\\\midrule
 & \multicolumn{4}{l}{\underline{Expected Exposure  }} & & \multicolumn{4}{l}{\underline{Value at Risk  }} & & \multicolumn{4}{l}{\underline{Expected Shortfall  }}\\

Scenario  & IRS & CDS & Two & Joint & & IRS & CDS & Two & Joint & & IRS & CDS & Two & Joint\\
 &   CCP & CCP & CCPs & CCP &  & CCP & CCP & CCPs & CCP & & CCP & CCP & CCPs & CCP\\\midrule
 \multicolumn{4}{l}{\underline{$t$-distributed CDS exposures}}\\
 JP Morgan Chase & 0.68 & 1.03 & 0.64 & 0.57 & & 0.94 & 1.01 & 0.92 & 0.90 & & 0.95 & 1.01 & 0.94 & 0.92\\
 Bank of America & 0.63 & 1.03 & 0.61 & 0.55 & & 0.93 & 1.01 & 0.91 & 0.90 & & 0.94 & 1.01 & 0.93 & 0.92\\
 Goldman Sachs & 0.72 & 1.02 & 0.62 & 0.53 & & 0.95 & 1.00 & 0.90 & 0.89 & & 0.96 & 1.00 & 0.92 & 0.91\\
 Morgan Stanley & 0.73 & 1.01 & 0.58 & 0.48 & & 0.95 & 1.00 & 0.88 & 0.87 & & 0.96 & 1.00 & 0.90 & 0.89\\
 Citigroup & 0.81 & 1.03 & 0.78 & 0.71 & & 0.97 & 1.01 & 0.95 & 0.94 & & 0.98 & 1.00 & 0.96 & 0.95\\
 Wells Fargo & 0.75 & 1.03 & 0.75 & 0.70 & & 0.95 & 1.01 & 0.95 & 0.93 & & 0.96 & 1.01 & 0.96 & 0.95\\
 HSBC & 0.95 & 0.89 & 0.68 & 0.57 & & 0.99 & 0.96 & 0.89 & 0.88 & & 0.99 & 0.96 & 0.91 & 0.89\\
 Taunus & 1.03 & 0.99 & 1.01 & 0.97 & & 1.01 & 0.99 & 1.00 & 0.98 & & 1.01 & 0.99 & 1.00 & 0.98\\
 Bank of New York & 0.95 & 1.00 & 0.95 & 0.95 & & 0.99 & 1.00 & 0.99 & 0.99 & & 0.99 & 1.00 & 0.99 & 0.99\\
 State Street & 1.01 & 1.00 & 1.01 & 1.01 & & 1.00 & 1.00 & 1.00 & 1.00 & & 1.00 & 1.00 & 1.00 & 1.00\\\midrule
 Total & 0.70 & 1.03 & 0.64 & 0.56\\\bottomrule
\end{tabular}\label{riskmeasures2009rhoZero.table}
}
}
\end{table}

\begin{table}[htbp] \centering
\caption{Reduction  of  interdealer exposures in various clearing scenarios, relative to the base scenario without central clearing, based on OCC data for gross  notional sizes (Q1 2009), assuming
 correlated exposures  across asset classes with $\rho=0.1$.
Top: normally distributed CDS exposures.  Bottom: $t$-distributed CDS exposures.}\smallskip
\makebox[\linewidth]{
\footnotesize{
\begin{tabular}
[c]{lrrrrrrrrrrrrrr}\toprule
 & \multicolumn{4}{l}{\underline{Expected Exposure  }} & & \multicolumn{4}{l}{\underline{Value at Risk  }} & & \multicolumn{4}{l}{\underline{Expected Shortfall  }}\\
Scenario  & IRS & CDS & Two & Joint & & IRS & CDS & Two & Joint & & IRS & CDS & Two & Joint\\
 &   CCP & CCP & CCPs & CCP &  & CCP & CCP & CCPs & CCP & & CCP & CCP & CCPs & CCP\\\midrule
 \multicolumn{4}{l}{\underline{Gaussian CDS exposures}}\\
 JP Morgan Chase & 0.71 & 1.00 & 0.63 & 0.56 & & 0.94 & 1.00 & 0.91 & 0.90 & & 0.96 & 1.00 & 0.93 & 0.92\\
 Bank of America & 0.66 & 1.01 & 0.60 & 0.54 & & 0.93 & 1.00 & 0.90 & 0.89 & & 0.94 & 1.00 & 0.92 & 0.91\\
 Goldman Sachs & 0.76 & 0.98 & 0.60 & 0.52 & & 0.95 & 0.99 & 0.89 & 0.88 & & 0.97 & 0.99 & 0.91 & 0.91\\
 Morgan Stanley & 0.77 & 0.96 & 0.56 & 0.47 & & 0.96 & 0.98 & 0.88 & 0.87 & & 0.97 & 0.99 & 0.90 & 0.90\\
 Citigroup & 0.82 & 0.99 & 0.74 & 0.67 & & 0.96 & 0.99 & 0.93 & 0.92 & & 0.97 & 0.99 & 0.95 & 0.94\\
 Wells Fargo & 0.75 & 1.01 & 0.72 & 0.68 & & 0.94 & 1.00 & 0.93 & 0.92 & & 0.96 & 1.00 & 0.95 & 0.94\\
 HSBC & 0.96 & 0.80 & 0.62 & 0.53 & & 0.99 & 0.95 & 0.88 & 0.88 & & 0.99 & 0.96 & 0.91 & 0.90\\
 Taunus & 1.02 & 0.94 & 0.95 & 0.91 & & 1.00 & 0.97 & 0.97 & 0.96 & & 1.00 & 0.98 & 0.98 & 0.97\\
 Bank of New York & 0.91 & 1.00 & 0.91 & 0.91 & & 0.97 & 1.00 & 0.97 & 0.97 & & 0.98 & 1.00 & 0.98 & 0.98\\
 State Street & 1.01 & 1.00 & 1.01 & 1.01 & & 1.00 & 1.00 & 1.00 & 1.00 & & 1.00 & 1.00 & 1.00 & 1.00\\\midrule
 Total & 0.73 & 0.99 & 0.62 & 0.55\\\midrule
 & \multicolumn{4}{l}{\underline{Expected Exposure  }} & & \multicolumn{4}{l}{\underline{Value at Risk  }} & & \multicolumn{4}{l}{\underline{Expected Shortfall  }}\\
Scenario  & IRS & CDS & Two & Joint & & IRS & CDS & Two & Joint & & IRS & CDS & Two & Joint\\
 &   CCP & CCP & CCPs & CCP &  & CCP & CCP & CCPs & CCP & & CCP & CCP & CCPs & CCP\\\midrule
 \multicolumn{4}{l}{\underline{$t$-distributed CDS exposures}}\\
 JP Morgan Chase & 0.68 & 1.01 & 0.63 & 0.57 & & 0.93 & 1.00 & 0.91 & 0.90 & & 0.94 & 1.00 & 0.92 & 0.92\\
 Bank of America & 0.63 & 1.01 & 0.60 & 0.55 & & 0.92 & 1.00 & 0.90 & 0.89 & & 0.93 & 1.00 & 0.92 & 0.91\\
 Goldman Sachs & 0.71 & 0.99 & 0.61 & 0.53 & & 0.93 & 0.99 & 0.89 & 0.88 & & 0.95 & 0.99 & 0.91 & 0.90\\
 Morgan Stanley & 0.71 & 0.98 & 0.56 & 0.48 & & 0.94 & 0.99 & 0.88 & 0.87 & & 0.95 & 0.99 & 0.90 & 0.89\\
 Citigroup & 0.79 & 1.00 & 0.75 & 0.68 & & 0.95 & 0.99 & 0.93 & 0.92 & & 0.96 & 0.99 & 0.94 & 0.94\\
 Wells Fargo & 0.74 & 1.01 & 0.72 & 0.68 & & 0.94 & 1.00 & 0.93 & 0.92 & & 0.95 & 1.00 & 0.95 & 0.94\\
 HSBC & 0.91 & 0.86 & 0.65 & 0.56 & & 0.97 & 0.95 & 0.88 & 0.87 & & 0.98 & 0.95 & 0.90 & 0.89\\
 Taunus & 1.01 & 0.96 & 0.97 & 0.93 & & 1.00 & 0.98 & 0.98 & 0.97 & & 1.00 & 0.98 & 0.98 & 0.97\\
 Bank of New York & 0.91 & 1.00 & 0.91 & 0.91 & & 0.97 & 1.00 & 0.97 & 0.97 & & 0.98 & 1.00 & 0.98 & 0.98\\
 State Street & 1.01 & 1.00 & 1.01 & 1.01 & & 1.00 & 1.00 & 1.00 & 1.00 & & 1.00 & 1.00 & 1.00 & 1.00\\\midrule
 Total & 0.70 & 1.00 & 0.62 & 0.56\\\bottomrule
\end{tabular}\label{riskmeasures2009rho01.table}
}
}
\end{table}

Tables \ref{riskmeasures2009rhoZero.table} and \ref{riskmeasures2009rho01.table} report the results
for the cases where the exposures are decorrelated and correlated with $\rho=0.1$, respectively.
From Table \ref{riskmeasures2009rhoZero.table}, comparing the case of no central clearing
in the market to the case where the OTC interest
rate swaps are cleared (columns 1, 5 and 9), we see that the overall exposures are reduced
significantly for almost all dealers and in all risk measures. The notional positions of the
dealers who have no reduction are very small compared to the total notionals of the
10 biggest dealers.
The rows with totals report the total expected exposure of all 20 dealers
relative to the total expected exposures in the scenario 0 with no CCP. We only compute the
``total" for the expected exposures as the other two risk measures are not additive.

While it is debatable whether the introduction of a clearinghouse for CDSs will reduce exposures when
there is no existing clearinghouse in the market (columns 2, 6 and 10), the results are more definitive
when the interest rate
class is already being cleared (columns 3, 7 and 11).
For both distributional choices of the CDS class, adding a CCP for the CDS class decreases the three
different risk measures. Depending on the assumptions made, the total expected exposures are
decreased with additional 6-11\% compared to the case when only interest rate swaps are cleared.
In absolute values this corresponds to reductions in the order of USD 41-76 billions in total expected net exposure.

In  the case where a single CCP clears both IRSs and CDSs the net exposure reductions are
higher since multilateral netting is possible across both types of derivatives classes. We also note
that the difference in exposure reductions between scenarios 1) and 3) are bigger than the difference between scenarios
3) and 4), and especially for the risk measures VaR and expected shortfall. Hence, the decision to clear
the CDS asset class is the most important, whether it is done by the same CCP that clears IRSs or not.

As expected, the results reported in Table \ref{riskmeasures2009rho01.table} confirm that in
the presence of positive asset class correlation, introduction of central clearinghouses increases the benefit
from clearing compared to the zero correlation case. All the fractions are slightly smaller with
a correlation parameter of $\rho=0.1$. When the asset classes are positively correlated the exposure reductions due to bilateral
netting across classes are reduced.

Interestingly, we note that using $t$-distributed CDS  exposures instead of Gaussian exposures does not change the conclusions.

For the different scenarios, Table \ref{maxexposures.table} reports the average of the maximum realized net exposure across dealers, which can be interpreted as
a worst case measure of dealer exposure. Yet again we observe that scenarios 1),3), and 4) result in significant exposure reductions.

\begin{table}[htbp] \centering
\caption{Average of the maximum exposure across counterparties in the
five scenarios and for the four different exposure specifications. All the numbers are
in millions.}\smallskip
\begin{tabular}
[c]{lrrrrr}\toprule
Scenario & Bilateral & IRS & CDS & Two & Joint\\
 &  & CCP & CCP & CCPs & CCP\\\midrule
 Gaussian CDS, $\rho=0$ & 136,460 & 114,340 & 138,830 & 108,110 & 103,190\\
 Gaussian CDS, $\rho=0.1$ & 145,040 & 120,440 & 144,850 & 113,130 & 109,160\\
 t-distributed CDS, $\rho=0$ & 136,020 & 111,630 & 138,480 & 107,770 & 103,070\\
 t-distributed CDS, $\rho=0.1$ & 144,140 & 117,500 & 144,370 & 112,650 & 108,820\\\bottomrule
\end{tabular}\label{maxexposures.table}
\end{table}

\begin{figure}[h]
  \begin{center}
    \includegraphics[width=1\textwidth]{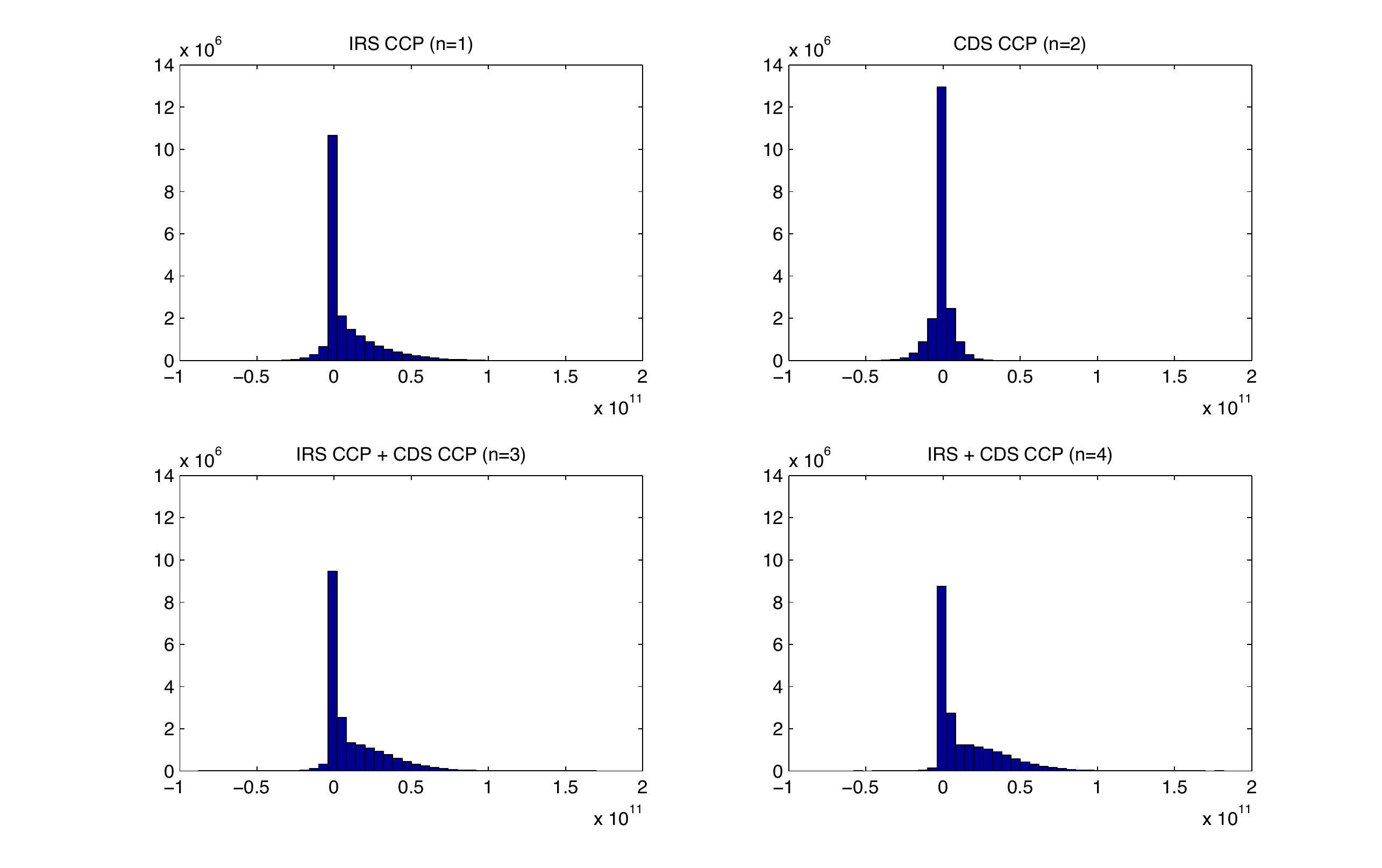}
    \caption{Histograms of the simulated differences \eqref{exposdiff} in the Gaussian CDS exposure zero correlation specification.
    Here all the exposure differences of the CCP participants are bundled together in a single vector.}
    \label{Histograms_fig}
  \end{center}
\end{figure}

For the different clearinghouse scenarios, Figure \ref{Histograms_fig} depicts the histograms of the reduction in exposures across counterparties
\begin{equation}
\epsilon_i^n=e_i^0-e_i^n,
\label{exposdiff}
\end{equation}
where $n=1,...,4$ corresponds to the clearing scenario,  in a market with 20 participants and  independent Gaussian exposures.
Positive values correspond to cases where central clearing reduces the exposure. Hence, scenarios
1),3), and 4) are generally giving rise to positive reductions in exposures. In particular it is observed how the CDS clearing
scenarios 3) and 4)
significantly reduce the tail of the distribution of differences compared to scenario 1) where only IRS are cleared. 

\newpage
\section{Conclusion}\label{conclusion_sec}

Using a stylized model  of  OTC exposures with multiple asset classes, we have compared
 the effects of bilateral netting across classes with those of multilateral netting across counterparties as allowed in a CCP.
 We find that the impact, on total expected exposure, of introducing a central
clearing facility for a given class of OTC derivatives is highly sensitive to  assumptions
of heterogeneity across asset classes in terms of 'riskyness' ,
as measured by the exposure per dollar
notional, as well as  correlation of exposures across asset classes. 
The base example
(IID, homogeneous Gaussian exposures) previously considered by \cite{duffiedoes} suggested that CCPs are inefficient in
reducing exposures unless one has an unrealistically high number of
participants, but the opposite conclusion is reached if plausible model parameters
are used to take into account differences in
riskyness across asset classes.
A second finding is that, when a CCP already exists for interest rate swaps, clearing of credit derivatives
decreases  exposures whether it is done by a new CCP or by the same CCP clearing IRSs.
This result distinguishes our study from \cite{duffiedoes} where the
marginal effect of clearing credit derivatives is very limited when interest rate swaps are
already cleared.
The findings are robust to distributional assumptions on (CDS) exposures.

We find, as in \cite{duffiedoes}, that the exposure reduction is highest if one CCP
clears all asset classes. However, this scenario would lead to a high concentration of systemic risk in the clearinghouse and also expose it to a high level of operational risk,   since
simultaneous clearing of different asset classes
requires a more sophisticated risk management technique. A serious examination of the benefits and drawbacks  of having multiple CCPs  cannot be done solely based on expected exposures and requires a model where systemic risk can be quantified as part of the tradeoff.

\newpage



\end{document}